\documentclass[12pt]{article}
\usepackage{amsmath}
\usepackage{amssymb}
\usepackage{rotating}
\usepackage{graphicx}
\usepackage{citesort}
\begin{document}

\begin{flushright}
{\small BARI-TH 405/2001} \\
\end{flushright}

\def\obs{{\cal O}}

\thispagestyle{empty}

\begin{center}
{\Large \bf {Lattice measurement of the scalar propagator}} \\[.6cm] 
{\Large \bf {near the symmetry breaking phase transition}}
\end{center}

\vspace{1.0cm}
\begin{center}
{\large 
P. Cea$^{1,2}$,
M. Consoli$^{3}$,
L. Cosmai$^{2}$ \\
\vspace{1.0cm}
{\small
$^1$~Dipartimento di Fisica,  Universit\`a di Bari,  via Amendola 173, 
I 70126 Bari, Italy\\[0.05cm] 
$^2$~INFN - Sezione di Bari, via Amendola 173, I 70126 Bari, Italy \\[0.05cm]
$^3$~INFN - Sezione di Catania, Corso Italia 57, I 95129 Catania, Italy 
\\[0.05cm] } }

\end{center}
\vspace{0.8cm}
\vspace{0.5cm}
\begin{center}
January, 2001
\end{center}
\vspace{0.5cm}
\begin{center}
{\large {\bf Abstract}}
\end{center}
\vspace{0.3cm}
Recent lattice simulations of 
$(\lambda \Phi^4)_4$ theories in the broken phase show that :
a) the shifted field propagator is well reproduced by the simple 2-parameter
form ${{Z_{\rm prop} }\over{ p^2 + M^2_h}}$ at finite momenta but strongly
differs for $p \to 0$~  b) the
bare zero-momentum two-point function 
$ \Gamma_2(0)= 
\left. \frac{ d^2 V_{\rm eff}}{d \varphi^2_B} \right|_{\varphi_B= \pm v_B} $
gives a value
of $Z_\varphi \equiv {{M^2_h}\over{\Gamma_2(0)}}$ that
increases when approaching the continuum limit. This supports theoretical
expectations where $v_B$ is related by an infinite re-scaling to the
 `physical Higgs condensate' 
$v_R$ defined through
$\left. \frac{ d^2 V_{\rm eff}}{d \varphi^2_R} \right|_{\varphi_R= \pm v_R}=M^2_h $.
New lattice data collected around  the phase transition 
confirm this scenario. By denoting
$M_{\rm SB} \equiv M_h ={\cal O} (v_R)$ the scale
of the broken phase, our results suggest
the existence of a `hierarchy' of scales
$\Gamma_2(0) \ll M^2_{\rm SB} \ll v^2_B$ that become infinitely far in the
continuum limit. This may open unexpected possibilities to reconcile an 
infinitesimal slope of the effective potential with finite values of $M_h$ and
accomodate very different mass scales in the framework of a spontaneously 
broken theory.

\newpage
\setcounter{page}{1}

\section{Introduction}

Spontaneous symmetry breaking through an elementary 
scalar field is the basic ingredient for the origin of particle 
masses in the Standard Model of electroweak interactions. Traditionally, 
the `condensation' of a scalar field, i.e. the transition 
from a symmetric phase where $\langle \Phi \rangle=0 $
to the physical vacuum where
$\langle \Phi \rangle \neq 0  $, has been described as an
essentially classical phenomenon in terms of
a classical potential (`B=Bare', $\lambda >0$)
\begin{equation}
\label{clpot}
   V_{\rm cl}(\Phi_B) =
 {{1}\over{2}} r_o \Phi^2_B + 
{{\lambda}\over{4!}}\Phi^4_B  
\end{equation}
with non-trivial absolute minima for a constant value $\Phi_B=\pm v_B \neq 0$. 
In this picture, 
by expanding around the absolute minima of the classical 
potential say $\Phi_B(x) = v_B +h(x) $, one predicts a simple relation 
\begin{equation}
\label{Mh}
           M^2_h= V''_{\rm cl}(\Phi_B = v_B)
\end{equation}
between the `Higgs mass' $M_h$ and the quadratic shape of the potential at the
minima. 

In the quantum theory, and on the basis of perturbation theory, Eq.(\ref{Mh})
is believed to represent a good approximation
by simply replacing the classical potential with the quantum
effective potential $V_{\rm eff}$. However, beyond perturbation theory, 
there is an alternative description of spontaneous symmetry breaking
\cite{zeit,agodi,plb,mech} 
where
$M^2_h$ and the curvature of the effective potential at the non-trivial minima are 
{\it different} physical 
quantities related by an infinite renormalization in the continuum limit 
of quantum field theory, with potentially important consequences for 
particle physics and cosmology.
The above  result has been obtained from gaussian 
and post-gaussian\cite{rit2} approximations to the effective potential. 
In view of the relevance of the issue and for the 
convenience of the reader, we shall
briefly recapitulate the main result in the simplest case
of the gaussian approximation to the energy density
${\cal E}_G[\varphi_B,\Omega]$ \cite{schiff,barnes,steve,cian,return}. This is
obtained from the
expectation value of the hamiltonian in the class of trial states with a constant
$\langle \Phi_B \rangle= \varphi_B$ and shifted-field
euclidean propagator 
\begin{equation}
\label{prop}
           G(p)= { { 1 }\over{ p^2 + \Omega^2 } }
\end{equation}
with a variational mass parameter $\Omega$. By minimization one gets
the coupled equations
\begin{equation}
\label{gap}
\Omega^2=r_o + {{\lambda \varphi^2_B}\over{2}} + 
{{\lambda}\over{2}} I_o(\Omega)
\end{equation}
and 
\begin{equation}
\label{tad}
0=r_o\varphi_B + {{\lambda\varphi^3_B}\over{6}}+
{{\lambda\varphi_B}\over{2}} I_o(\Omega)
\end{equation}
with
\begin{equation}
I_o(\Omega)=\int {{d^4p}\over{(2\pi)^4}} {{1}\over{p^2 + \Omega^2}}
\end{equation}
By using Eq.(\ref{gap}), one can define a $\varphi_B-$dependent mass 
$\Omega=\Omega(\varphi_B)$ and 
the gaussian approximation to the effective potential
$V_G(\varphi_B)= {\cal E}_G[\varphi_B,\Omega(\varphi_B)]$ whose first derivative 
has the simple form \cite{zeit}
\begin{equation}
\label{derivative}
 \frac{ d V_ G }{d \varphi_B} = \varphi_B [\Omega^2(\varphi_B) - 
{{\lambda \varphi^2_B}\over{3}}]
\end{equation}
In this way, spontaneous symmetry breaking
is associated with those absolute minima
$\varphi_B=\pm v_B \neq 0$ where 
\begin{equation}
\label{abs}
\Omega^2(v_B)= {{\lambda v^2_B}\over{3}} \equiv M^2_h
\end{equation}
Now, the zero-momentum 
two-point function (the inverse susceptibility)
defines the quadratic shape of the potential at the 
non-trivial minima 
\begin{equation}
\label{chichi}
 \chi^{-1}=  \Gamma_2(p=0)= 
\left. \frac{ d^2 V_{\rm eff} }{d \varphi^2_B} \right|_{\varphi_B= \pm v_B} \equiv
{{M^2_h}\over{Z_\varphi }}
\end{equation}
where we have introduced a 
re-scaling factor $Z=Z_\varphi$ {\it defined} through Eq.(\ref{chichi})
\cite{huang}. In the gaussian approximation, by using the identities \cite{steve}
\begin{equation}
I_o(0)-I_o(\Omega)= {{\Omega^2}\over{2}}[I_{-1}(\Omega) + {{1}\over{8\pi^2}}]
\end{equation}
and the definition
\begin{equation}
       {{dI_o}\over{d\Omega}}\equiv - \Omega I_{-1}(\Omega)
\end{equation}
we obtain 
\begin{equation}
 \frac{ d^2 V_ G }{d \varphi^2_B} =  \Omega^2(\varphi_B) - 
{{\lambda\varphi^2_B }\over{4}}~
{{  \lambda I_{-1}(\Omega)  }\over{ 1
   + {{\lambda I_{-1}(\Omega)}\over{4}}       }}
\end{equation}
In the symmetric vacuum where $\varphi_B=0$, one finds
\begin{equation}
\label{equal}
\left. \frac{ d^2 V_G }{d \varphi^2_B} \right|_{\varphi_B= 0}=
\Omega^2(0)
\end{equation}
while, in the broken-symmetry phase, the alternative expression
\begin{equation}
\label{Gammag}
\left. \frac{ d^2 V_G }{d \varphi^2_B} \right|_{\varphi_B= \pm v_B}=
 M^2_h~ {
{1 - {{\lambda}\over{2}}I_{-1}(M_h)}\over
{1 + {{\lambda}\over{4}}I_{-1}(M_h)}}
\end{equation}
For an evaluation of
Eq.(\ref{Gammag}), it is convenient
to re-define the bare mass term as
\begin{equation}
      r_o= -{{\lambda}\over{2}} I_o(0) + \Delta
\end{equation}
so that the gap-equation at the absolute minima reduces to
\begin{equation}
\label{gap2}
0= \Delta+ {{M^2_h}\over{2}} 
-{{\lambda M^2_h}\over{4}}[I_{-1}(M_h)+ {{1}\over{8\pi^2}}]  
\end{equation}
( $\Delta=0$ corresponds to the 
`Coleman-Weinberg regime ' \cite{CW} 
where $\Omega(0)=0$). By defining
\begin{equation}
          I_{-1}(M_h)\equiv {{1}\over{8\pi^2}} \ln {{\Lambda^2}\over{M^2_h}}
\end{equation}
and
 $\mu^2\equiv \Lambda^2 e^{-{{16\pi^2}\over{\lambda}} }$,~
    $ \Delta \equiv {{\lambda \mu^2}\over{32\pi^2}}y$,~  
        $M^2_h\equiv \omega^2 \mu^2$,~ Eq.(\ref{gap2}) becomes 
\begin{equation}
       \omega^2\ln {{e}\over{\omega^2}}= y 
\end{equation}         
Finally, by replacing into Eq.(\ref{Gammag}), we obtain the gaussian 
approximation prediction for $Z_\varphi$ in the broken phase
\begin{equation}
  {{1}\over{Z_\varphi}}= {{\Gamma_2(p=0)}\over{M^2_h}}= 
{ {  {{\lambda}\over{16\pi^2}} \ln\omega^2 } \over{  {{3}\over{2}}- 
    {{\lambda}\over{32\pi^2}} \ln\omega^2 } }
\end{equation}
that (in the range 
$ e^{ {{48\pi^2}\over{\lambda}} } >  \omega^2 >1$) amounts to
\begin{equation}
\label{zetag}
 Z_\varphi= {{24\pi^2}\over{ \lambda \ln \omega^2}}-{{1}\over{2}}
\end{equation}
Eq.(\ref{zetag}) 
can be considered from two qualitatively different points of view.
On one hand, 
for given values of the dimensionless parameters $y$ and $\omega^2$, it
implies a behaviour $Z_\varphi \sim {{1}\over{\lambda}}$. Thus, in a
continuum limit where $\Lambda \to \infty$ and $\mu^2$ (and hence
$M^2_h=\mu^2\omega^2$) is kept fixed, one gets
$Z_\varphi\sim \ln {{\Lambda^2}\over{M^2_h}} \to \infty$ implying that
the `Higgs mass' $M_h$ and the curvature of the effective potential are 
{\it different} physical quantities. Indeed, they would be
related by an infinite renormalization in the
continuum limit of quantum field theory.

On the other hand, for any given $\lambda$, i.e.
when $\mu^2$ and $\Lambda$ are kept in a fixed ratio, 
Eq.(\ref{zetag}) predicts that $Z_\varphi$ should
also become larger while decreasing $\omega^2$. This corresponds to increase
$y$ from negative values trying to
approach the minimum value $\omega^2=1$. However, in this latter case, 
$Z_\varphi$ will not diverge. In fact, before reaching
the value $\omega^2=1$, the gaussian effective potential exhibits a (weakly)
first-order phase transition to the symmetric phase \cite{zeit}.

We stress that the gaussian-approximation
prediction of a non trivial $Z_\varphi$ 
has been tested  with precise lattice simulations 
\cite{cea1,cea2} performed in the Ising limit of the theory.
The data show that by approaching the critical value of the hopping parameter
$\kappa_c\sim 0.0748$ \cite{montweisz} from the broken phase, 
the quantity $Z_\varphi\equiv M^2_h \chi$ rapidly increases above unity.
Thus, in the broken phase, $Z_\varphi$
 is different from the more conventional
quantity $Z=Z_{\rm prop}$ associated with the
two-parameter form for the
shifted-field propagator
\begin{equation}
\label{zprop}
G_{\rm pole}(p)= \frac{Z_{\rm prop}}{p^2 + M^2_h} ,
\end{equation}
According to `triviality' \cite{book}, 
this latter quantity should exhibit a continuum limit $Z_{\rm prop}\to 1$ for
consistency  with the K\"allen-Lehmann decomposition.
The point is that, in the broken phase, 
the data deviate from Eq.(\ref{zprop})
when $p \to 0$ 
presaging an even more dramatic 
difference at $p=0$. Furthermore, 
the discrepancy between $Z_\varphi$ and $Z_{\rm prop}$
becomes {\it larger} in the limit 
$\kappa \to \kappa_c$. Therefore, it cannot be explained by introducing residual
perturbative corrections that, according to `triviality'\cite{book}, should
become smaller and smaller in the same limit.
On the other hand, in the symmetric phase where
Eq.(\ref{zprop}) describes the lattice data down to $p=0$
\cite{cea2}, one gets $Z_\varphi=Z_{\rm prop}\sim 1$ 
as expected on the basis of Eq.(\ref{equal}). 

After this general Introduction, we shall report in this Letter
further numerical evidence for the
validity of this theoretical framework
by studying the $\omega-$dependence of
Eq.(\ref{zetag}). Strictly speaking, to this end, one should work at fixed 
$\lambda$ by changing the bare mass in the full
two-parameter $\lambda\Phi^4$ theory, the Ising limit being just
a one-parameter model. However, even in the 
Ising limit we can reproduce the situation of a two-parameter theory if we
perform a finite-temperature simulation. Namely, 
by changing the value of $\kappa$ in the broken phase, we know from 
refs.\cite{cea1,cea2} that the
zero-temperature value of $Z_\varphi$ increases when $\kappa \to \kappa_c$.
However, for any fixed value $\kappa > \kappa_c$,  by increasing the 
temperature, we shall approach  the phase transition. In this case, we expect
the lattice data to exhibit a qualitatively different 
increase of $Z_\varphi$, analogously to the effect induced in Eq.(\ref{zetag})
by approaching the phase transition at fixed $\lambda$
in the full two-parameter theory.

\section{Finite-temperature lattice simulation}

The transition to finite-temperature field theory 
is easily performed by following the formalism introduced in 
refs.\cite{bernard}. In the case of bosonic fields the
partition function is
\begin{equation}
\label{partition}
    {\cal Z}(\beta)= \int {\cal D}\Phi~e^{-\int^{\beta}_{0} d\tau \int 
d^3{\bf{x}} {\cal L}_E}
\end{equation}
where $\beta=1/T$ is the inverse temperature, ${\cal{L}}_E $ is the Euclidean lagrangian
density and the functional integral 
is performed over all periodic field configurations with 
$\Phi({\bf{x}},\tau=0)=\Phi({\bf{x}},\tau=\beta)$. 
In Fourier space, the compactification of the time direction leads to a
discrete energy spectrum with energies $\omega_n={{2n\pi}\over{\beta}}$ and,
therefore, to the form of a free propagator 
\begin{equation}
\label{gfree}
G^{(\beta)}_{\rm free}(n, {\bf{ p}} )= 
{ { 1 }\over 
{   \hat { {\bf{ p}} }^2 + \hat {  p_4 }^2 + M^2 } }
\end{equation}

As it is well known~\cite{montmunster}, a Monte-Carlo
simulation with periodic boundary conditions
on an asymmetric $L^3_s ~{\rm x}~L_t$  lattice 
is equivalent to a finite-temperature calculation at $\beta=L_t=1/T$. 
To this end, a one-component 
$(\lambda\Phi^4)_4$ theory   
\begin{equation}
\label{action}
   S =\sum_x \left[ \frac{1}{2}\sum_{\mu}(\Phi(x+\hat e_{\mu}) - 
\Phi(x))^2 + \frac{r_0}{2}\Phi^2(x)  + \frac{\lambda_0}{4} \Phi^4(x)  
\right]    
\end{equation}
is conveniently studied in the Ising limit 
\begin{equation}
\label{ising}
   S_{\rm Ising} = -\kappa
\sum_x\sum_{\mu} \left[ 
\phi(x+\hat e_{\mu})\phi(x) +
\phi(x-\hat e_{\mu})\phi(x) \right]    
\end{equation}
with $\Phi(x)=\sqrt{2\kappa}\phi(x)$ and where $\phi(x)$ takes only the 
values $+1$ or $-1$.  

    To perform Monte-Carlo simulations of this Ising action, we have used
the Swendsen-Wang \cite{SW} cluster algorithm. Statistical errors 
can be estimated through a direct evaluation of the integrated autocorrelation 
time~\cite{Madras88}, or by using the ``blocking''~\cite{blocking} or the 
``grouped jackknife''~\cite{jackknife} algorithms.  We have checked that 
applying these three different methods we get consistent results.  
          Lattice observables include:

(i) the bare magnetization, 
$v_B=\langle |\Phi| \rangle$, 
where $\Phi \equiv \sum_x \Phi(x)/L^4$ is the average field for each 
lattice configuration. The broken phase is found for $\kappa > \kappa_c$
where $\kappa_c \simeq 0.0748$ \cite{montweisz} in an infinite volume.
For any $\kappa > \kappa_c$, we expect to detect a 
phase transition to the symmetric phase at some sufficiently small value 
of $L_t$. 

(ii) the zero-momentum susceptibility 
\begin{equation}
\label{suscep}
\chi=L^4 \left[ \left\langle |\Phi|^2 \right\rangle - 
\left\langle |\Phi| \right\rangle^2 \right] ,
\end{equation}

(iii) 
the shifted-field propagator 
\begin{equation}
\label{shifted}
G(p)= \left\langle \sum_x \exp (ip x) (\Phi(x) - v_B) 
(\Phi(0)- v_B) \right\rangle \, ,
\end{equation}
where $p\equiv( p_4,{\bf{p}})$ with
${\bf {p}}=\frac{2\pi}{L_s}{\bf{n}}$ 
and $p_4=\frac{2\pi}{L_t}{n_4}$ where $(n_1,n_2,n_3,n_4)$ are
integer-valued components, not all zero.

To extract
the `Higgs mass' $M_h$ one has to preliminarly compare
the lattice data for $G(p)$ with the 2-parameter formula 
\begin{equation}
\label{gbetaform}
G^{(\beta)}_{\rm fit}(n, {\bf{ p}} )= 
{ { Z_{\rm prop} }\over 
{   \hat { {\bf{ p}} }^2 + \hat {  p_4 }^2 + m^2_{\rm latt} } }
\end{equation}
where $m_{\rm latt}$ is the mass in lattice units and 
$\hat{p}_\mu= 2 \sin{{p_\mu}\over{2}}$.

Obviously, by setting $L_t=L_s$ we re-obtain the zero temperature case
studied in refs.\cite{cea1,cea2}. We now repeat the basic point of \cite{cea2}.
To realize how good the fit with Eq.(\ref{gbetaform}) can be, we start 
at $T=0$ in the symmetric phase at  $\kappa=0.0740$ on a $20^4$ lattice. 
In Fig.1, we report the same data of ref.\cite{cea2} where
the scalar propagator has been suitably re-scaled
to show the very good quality of the fit to Eq.~(\ref{gbetaform}). 
The value of $Z_{\rm prop}$ is indicated by the dashed line while
$Z_\varphi$ is reported in Fig.1 as a black point at $\hat{p}=0$.
Notice the perfect agreement between $Z_\varphi$ and  $Z_{\rm prop}$. 

We now select a value $\kappa=0.07512$, in the broken phase, where the 
2-parameter fit
to the propagator data yields a comparable value for $m_{\rm latt}$ by using
a $32^4$ lattice. However, 
unlike Fig.~1, the fit to Eq.~(\ref{gbetaform}), though excellent at higher 
momenta, does not reproduce the lattice data down to zero-momentum. 
Therefore, in the broken phase, a meaningful determination of $Z_{\rm prop}$ 
and $m_{\rm latt}$ requires excluding the lowest momentum points from the 
fit. The lattice data are shown in Fig.2. 

By fixing $\kappa=0.07512$, we have now performed 
computations on asymmetric $L^3_s \times  L_t$ 
lattices to simulate a finite-temperature system. 
Notice that well past the phase transition, i.e.
for very small $L_t$, consistency requires 
a $p \to 0$ limit of the propagator such that
$Z_\varphi$ 
and $Z_{\rm prop}$ agree to good accuracy.
In fact, if running on {\it smaller} lattices one is able
to re-obtain the same conditions
 of the symmetric-phase calculation in Fig.1, 
we can exclude the presence of finite-size artifacts in the 
zero-temperature simulation and interpret the discrepancy between 
$Z_\varphi$ and $Z_{\rm prop}$ as a real physical
effect due to the presence of the scalar condensate. At the same time, 
the phase transition associated with
spontaneous symmetry breaking can be
understood as a real condensation process where the
$Z_\varphi - Z_{\rm prop}$ discrepancy represents a distinctive
non-perturbative feature.

This expectation
can be checked in Fig.~3 by comparing 
the data taken on the $32^{3} \times 2$ lattice with those taken 
on the $32$-times bigger $64^{3} \times 8$ lattice when
the system, however, 
is still in the broken phase and a meaningful determination 
of $m_{\rm latt}$ through
Eq.(\ref{gbetaform}) requires excluding the lowest momentum points from
the fit.

The data 
for the physical observables from different lattice sizes are reported in
Table 1. For all lattice sizes there is a clear evidence for a 
phase transition in the region
$6 < L_t < 8$ where the system crosses from the broken into the symmetric phase
(see Fig.~4).

We stress that the conventional interpretation of `triviality', assuming
$Z_\varphi = Z_{\mathrm{prop}} \simeq 1$, predicts
that approaching the phase transition one finds 
$m^2 \to 0$ and  $\chi \to \infty$ in such a way that
$m^2 \chi$ remains constant. However, our data 
 for the quantity $Z_\varphi \equiv m^2_{\rm latt}\chi$, reported in Fig.~5,
 exhibit a clear increase in the  critical region while the corresponding value
for $Z_{\mathrm{prop}}$ remains remarkably constant (see Table 1).
This confirms once more that $Z_\varphi$, as defined in Eq.(\ref{chichi}), and
$Z_{\text{prop}}$, as defined in Eq.(\ref{zprop}), are different physical quantities.

\section{Summary and outlook}

The finite-temperature simulations presented in this paper confirm the 
results of refs.\cite{cea1,cea2}. Both provide 
strong numerical evidence that, close to the continuum limit of quantum field
theory, a
naive perturbative description 
of a spontaneously broken $(\lambda\Phi^4)_4$ is missing
very important effects.
Although the theory is `trivial'  there are 
non-perturbative
collective effects at (and near) zero momentum producing the observed large
deviations from the standard free-field-like form of the propagator
Eq.(\ref{gbetaform}). 
These deviations are responsible for the non-trivial
re-scaling factor 
$Z_\varphi=M^2_h \chi$ that increases rapidly when
$\kappa \to \kappa_c$ in the zero-temperature case and when approaching the
phase transition
in the finite-temperature simulations. This provides additional numerical evidence 
for the validity of Eq.(\ref{zetag}) that represents 
a distinctive prediction and
can be used, for instance, to reconcile a finite
value of $M_h$ with an
infinitesimal quadratic shape of the effective potential at its minima.
Indeed, with a divergent $Z_\varphi$, the usual definition of the
physical vacuum field $v_R$ 
\begin{equation}
\left. \frac{ d^2 V_{\rm eff}}{d \varphi^2_R} \right|_{\varphi_R= \pm v_R} = M^2_h
\end{equation}
is related by a non-trivial infinite
re-scaling to the bare `Higgs condensate' $v_B$ in Eq.(\ref{Gammag}).
Therefore, although $v_B$ 
diverges in units of $M_h$, one can obtain a continuum limit where {\it both}
$M_h$ and $v_R$ are  finite quantities setting the scale 
of the spontaneously broken phase $M_{\rm SB}\equiv M_h= {\cal O} (v_R)$.
 In this sense, a divergent $Z_\varphi$ means that
 spontaneous symmetry breaking introduces  a mass `hierarchy' 
\cite{mech} where $\Gamma_2(p=0) \ll M^2_{\rm SB} \ll v^2_B$ are  infinitely
far scales in the continuum limit since
\begin{equation}
\label{hierarchy}
 Z_\varphi  ={{M^2_h }\over{ \Gamma_2(p=0)}} = {{v^2_B}\over{v^2_R}} \sim 
\ln  {{\Lambda } \over{ M_h}} \to \infty
\end{equation}

Another interesting point concerns the actual form of the energy spectrum for the
long-wavelength excitations
of the broken phase. In fact, the
observed differences of the propagator
from Eq.(\ref{zprop}) for $p \to 0$ imply that
the energy spectrum of the broken phase 
$\tilde{E} ({\bf{p}})$ 
sizeably differs from
$\sqrt { {\bf{p}}^2 + M^2_h }$ 
when ${\bf{p}} \to 0$. In this respect, the results of ref.\cite{cea2} 
show that the spectrum
$\tilde{E} ({\bf{p}})$ approaches the form
$\sqrt { {\bf{p}}^2 + M^2_h }$ only at large $ {\bf{p}}^2$.
Also, $\tilde{E}(0) < M_h$ and their difference increases in the
continuum limit. Such a difference, detected by studying 
the time-slices of the connected correlator at various values of
${\bf{p}}$, has no 
counterpart in the symmetric phase. In this latter case, the form
$ E ({\bf{p}})=\sqrt { {\bf{p}}^2 + m^2 }$ is found \cite{cea2} to
reproduce the lattice data to high accuracy for all ${\bf{p}}$
down to ${\bf{p}} = 0$ so that, in this case, one gets
$E(0)=m$.
A particularly important question concerns the stability of the results for
the energy-gap
$\tilde{E}(0)$ in the broken phase and for the `Higgs mass'
$M_h$ controlling the higher-momentum behaviour of the propagator.
The results of \cite{cea2} for $\kappa=0.076$, the case 
studied by Jansen et al. \cite{jansen} on a $20^4$ lattice, show that the
value of $M_h$, extracted from the set of higher-momentum data where one gets
a good fit with Eq.(\ref{zprop}), is remarkably stable
for variations of the lattice size 
from $20^4$ to $32^4$. New preliminary data~\cite{cea3} 
show that the energy spectrum 
$\tilde{E}({\bf{p}})$ 
remains also stable, at least for not too small values of
${\bf{p}}^2$.
On the other hand, by increasing the lattice size up to $32^4$, 
our present data suggest a decrease of
 $\tilde{E}(0)$ that, therefore, differently from $M_h$, 
may represent an infrared-sensitive
quantity. A complete discussion of this point requires, however, more
statistics and will be presented in a forthcoming paper~\cite{cea3} .


\begin{table}
\begin{sideways}
\begin{tabular}{cccccccc}
lattice&     \#configs.&  $1/L_t$& $<|\Phi|>$& $m_{\mathrm{latt}}$& $\chi$&  $Z_\varphi$&  $Z_{\mathrm{prop}}$\\
\hline
$32^3\times2$&   700K&  0.5&       0.015980(42)&  0.494362(917)&  26.25(0.14)&    0.963675(6389)&   0.960179(1167)\\
$32^3\times3$&   835K&  0.333333&  0.024602(47)&  0.261363(681)&  93.28(0.34)&    0.957328(6074)&   0.955672(898)\\
$32^3\times4$&   650K&  0.25&      0.035326(81))&  0.180635(6834)& 253.60(1.09)&   1.243208(94220)&  0.959723(1649)\\
$32^3\times5$&   875K&  0.2&       0.050104(105)&  0.139061(5478)& 614.61(2.36)&   1.785661(140851)& 0.959713(1306)\\
$32^3\times6$&   725K&  0.166667&  0.071158(285)&  0.125006(6219)& 1376.14(9.58)&  3.230797(322248)& 0.959374(1373)\\
$32^3\times8$&   725K&  0.125&  0.117674(375)&  0.171012(7257)& 530.55(3.85)&  2.331141(198569)& 0.959329(1677)\\
$32^3\times32$&  400K&  0&         0.161778(130)&  0.20623(409)&  193.1(1.7)&    1.233876(501)&  0.9551(21)\\
\hline
$48^3\times3$&   700K&  0.333333&  0.013413(27)&  0.261399(594)& 93.65(0.35)&  0.961384(5645)& 0.955420(834)\\
$48^3\times4$&   460K&  0.25&      0.019187(47)&  0.165589(2391)& 255.35(1.19)& 1.051909(30770)& 0.956229(1257)\\
$48^3\times5$&   635K&  0.2&       0.027706(72)&  0.118999(2923)& 655.07(3.26)&  1.393679(68817)& 0.958973(1088)\\
$48^3\times6$&   400K&  0.166667&  0.044522(193)& 0.114195(6459)& 1919.71(15.16)&  3.761108(426501)& 0.960837(1577)\\
$48^3\times8$&   170K&  0.125&     0.116546(516)&  0.157074(7182)& 826.14(20.94)&  3.062311(290602)& 0.958064(2392)\\
$48^3\times9$&   185K&  0.111111&  0.136291(415)&  0.180302(6762)& 464.83(11.35)&  2.270273(179082)& 0.955628(2456)\\
\hline
$64^3\times2$&   390K&  0.5&       0.005626(18)&  0.494478(1463)& 26.14(0.17)&    0.9604047(8320)&  0.959913(3017)\\
$64^3\times3$&   250K&  0.333333&  0.008712(31)&  0.260004(681)&  93.52(0.74)&    0.9498651(9044)&  0.952527(1263)\\
$64^3\times4$&   480K&  0.25&      0.012439(33)&  0.163052(2724)& 259.16(4.95)&    1.035163(39846)&  0.954254(1341)\\
$64^3\times5$&   200K&  0.2&       0.017980(89)&  0.110235(4120)&  676.43(19.80)&    1.234948(99135)&  0.955591(1730)\\
$64^3\times6$&    90K&  0.166667&  0.029673(263)&  0.090056(9592)& 2105.82(24.60)& 2.565853(547407)& 0.957225(2888)\\
$64^3\times8$&   240K&  0.125&  0.118429(299)&  0.162401(7698) & 803.64(17.45)&    3.184376(309700)& 0.957547(2466)\\
$64^3\times9$&   245K&  0.111111&  0.136991(178)&  0.177897(4993)& 417.49(7.65)& 1.985071(117215)& 0.955239(2092)\\
$64^3\times10$&  150K&  0.1&       0.1462464(180)&  0.180102(4437)& 311.83(4.56)&   1.519625(78097)&  0.952393(2373)\\
$64^3\times14$&   80K&  0.0714286& 0.158104(137)&  0.193444(5602)& 221.91(3.73)&   1.247618(75245)&  0.948202(3055)\\[1.0cm]
\multicolumn{8}{c}{Table 1: Summary of the simulation runs for the 4d $\lambda \varphi^4$ model
at finite temperature (Ising limit, $\kappa=0.07512$).}
\end{tabular}
\end{sideways}
\end{table}

\clearpage
\begin{figure}[t]
\begin{center}
FIGURE 1
\end{center}
\label{Fig1}
\begin{center}
\includegraphics[clip,width=1.0\textwidth]{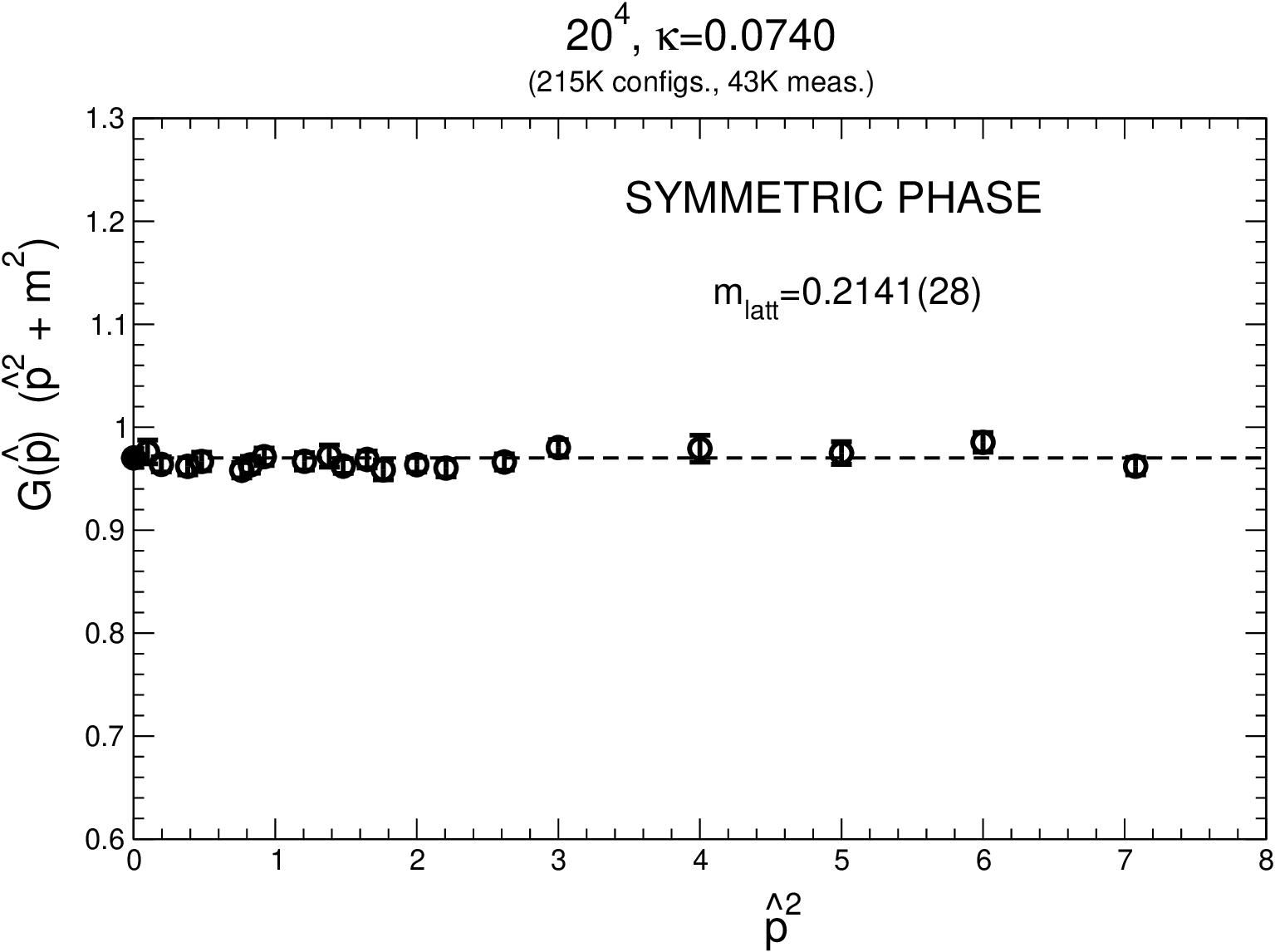}
\caption{The lattice data for the re-scaled propagator at $\kappa=0.0740$ in the
symmetric phase. The zero-momentum full point is defined as 
$Z_\varphi=m^2_{\mathrm{latt}}\chi$.
The dashed line indicates the value of $Z_{\mathrm{prop}}$.}
\end{center}
\end{figure}
\clearpage
\begin{figure}[t]
\begin{center}
FIGURE 2
\end{center}
\label{Fig2}
\begin{center}
\includegraphics[clip,width=1.0\textwidth]{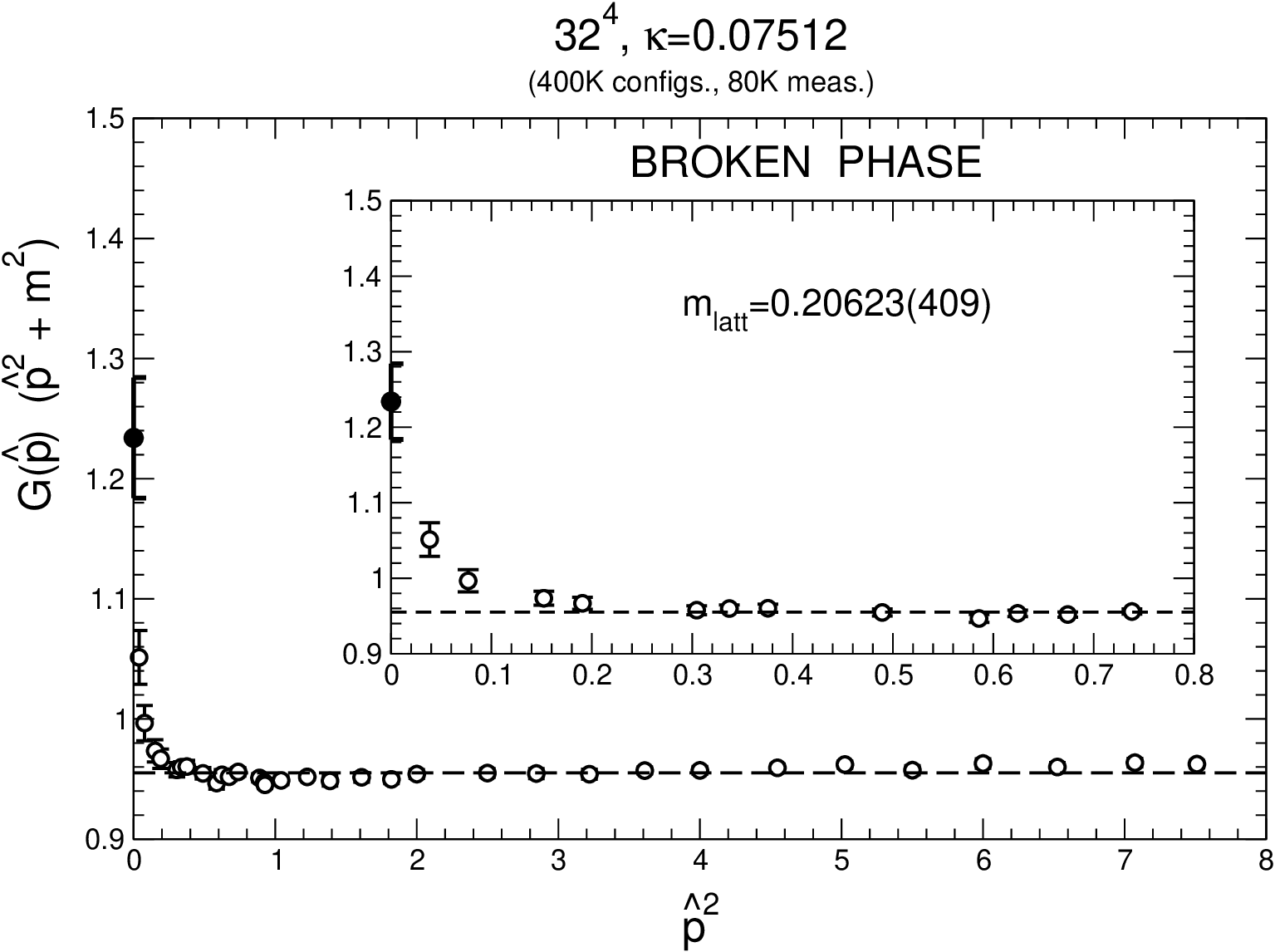}
\caption{The lattice data for the re-scaled propagator at $\kappa=0.07512$.
The zero-momentum full point is defined as $Z_\varphi=m^2_{\mathrm{latt}}\chi$.
The very low momentum region is shown in the inset. 
The dashed line indicates the value of $Z_{\mathrm{prop}}$.}
\end{center}
\end{figure}
\clearpage
\begin{figure}[t]
\begin{center}
FIGURE 3
\end{center}
\label{Fig3}
\begin{center}
\includegraphics[clip,width=1.0\textwidth]{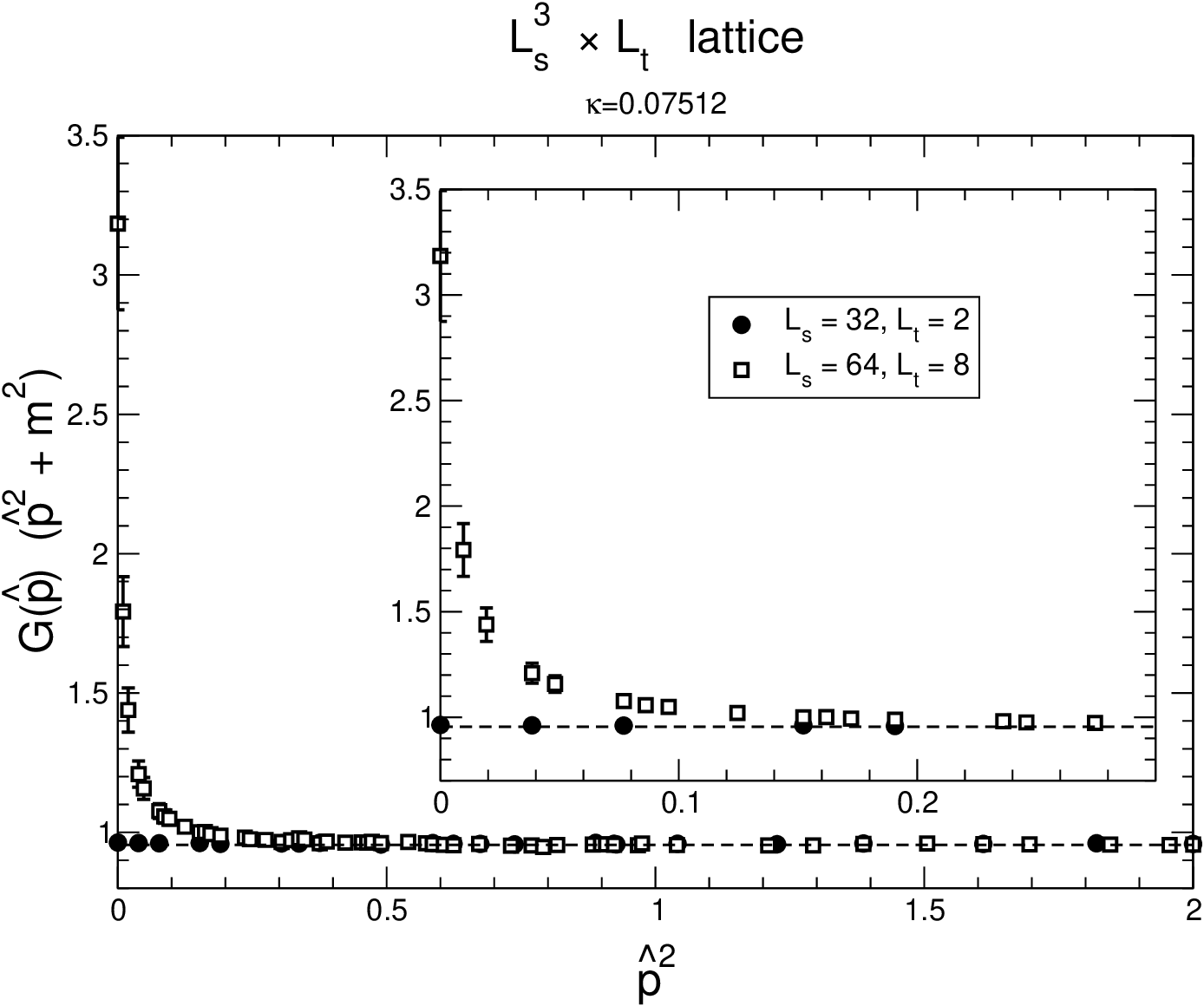}
\caption{The same as in Fig.~2 for $L_s=32$ and $L_t=2$,
and $L_s=64$ and $L_t=8$. The two sets of data are normalized
in terms of the corresponding lattice masses shown in Table 1.
The dashed lines indicate the same zero-temperature value
$Z_{\mathrm{prop}}=0.9551$ as in Fig.2}
\end{center}
\end{figure}
\clearpage
\begin{figure}[t]
\begin{center}
FIGURE 4
\end{center}
\label{Fig4}
\begin{center}
\includegraphics[clip,width=0.75\textwidth]{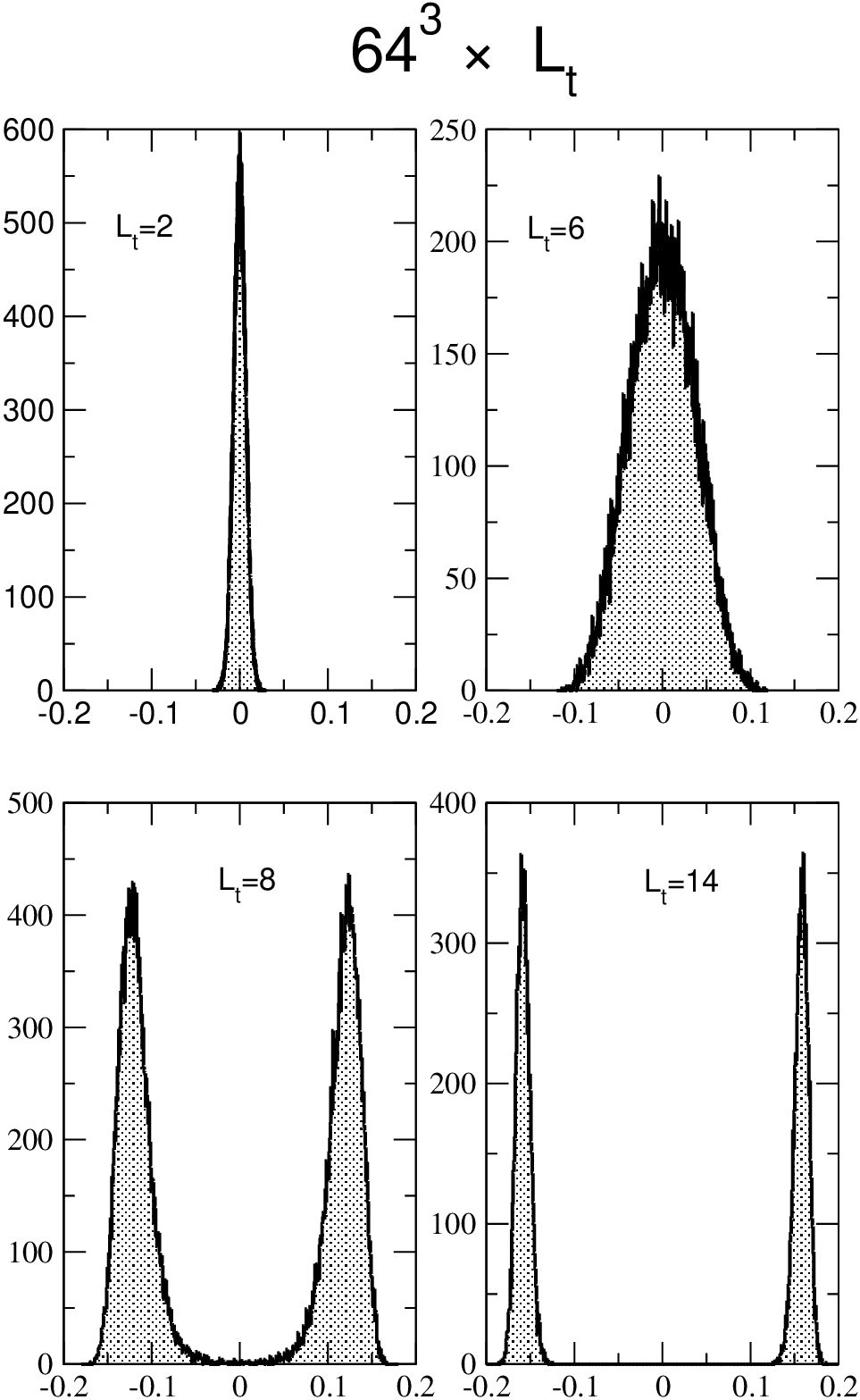}
\caption{The distribution of the average field value for 4 values
of $L_t$.}
\end{center}
\end{figure}
\clearpage
\begin{figure}[t]
\begin{center}
FIGURE 5
\end{center}
\label{Fig5}
\begin{center}
\includegraphics[clip,width=1.0\textwidth]{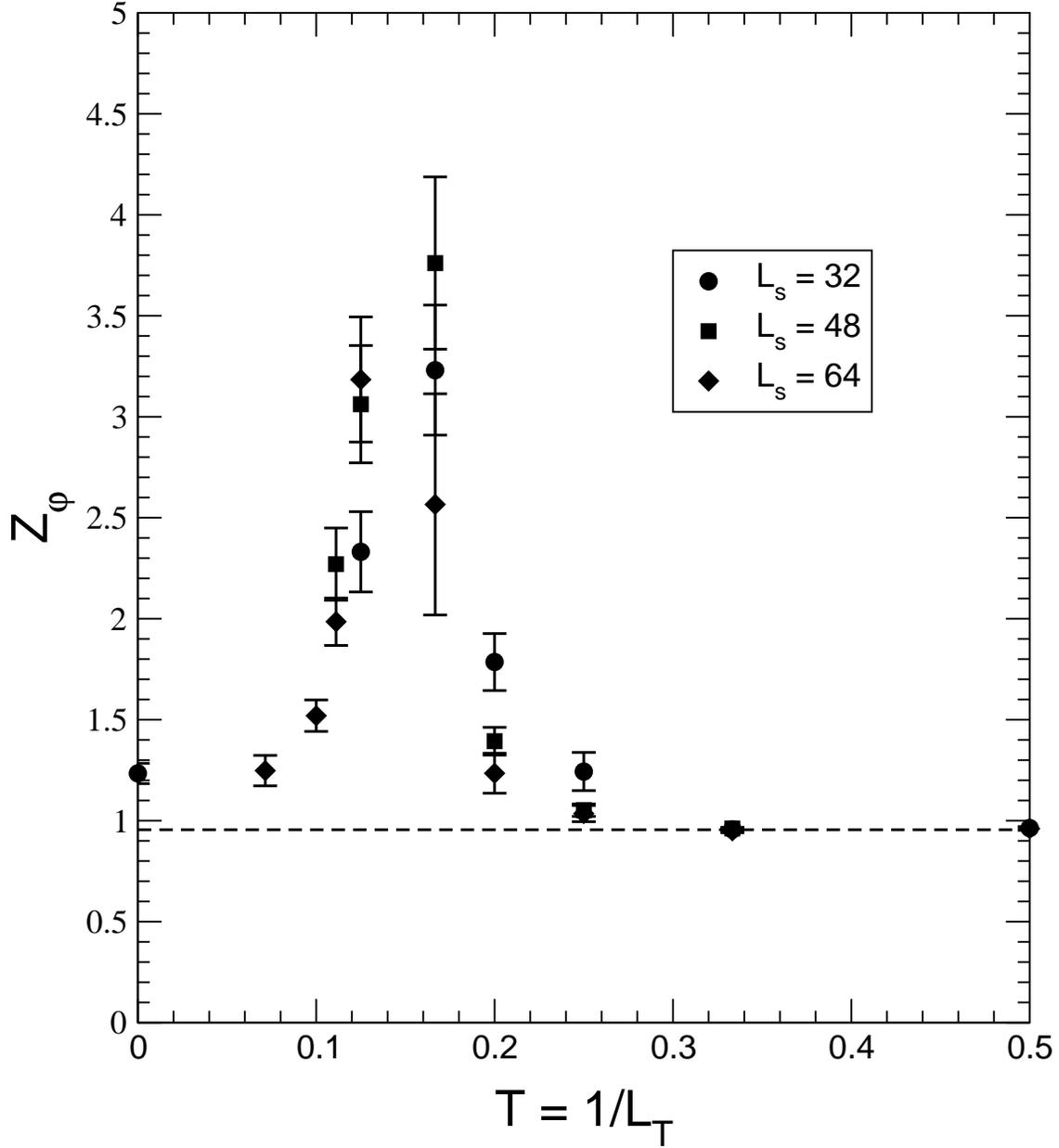}
\caption{The lattice data for $Z_\varphi \equiv m^2_{\mathrm{latt}} \chi$
for $L_s=32,48,64$  vs. the temperature.
The dashed line indicates the zero-temperature value
$Z_{\mathrm{prop}}=0.9551$ as in Fig.2}
\end{center}
\end{figure}
\clearpage

\end{document}